\documentclass[a4paper]{jpconf}  % for double-spaced preprint
\usepackage[dvips]{graphicx}
\usepackage{dcolumn}   % needed for some tables
\usepackage{bm}        % for math
\usepackage{amssymb}   % for math
\usepackage[utf8]{inputenc}
\usepackage{mathrsfs}
\usepackage{amsmath}
\usepackage[breaklinks=true,colorlinks=true]{hyperref}
\hypersetup{colorlinks=true,citecolor=blue,linkcolor=blue,urlcolor=blue}
\usepackage{textcomp}
\usepackage{subfigure}
\usepackage{color}

\begin{document}

\title{Domain wall thickness and deformations of the field model}

\author{Petr A. Blinov$^{1}$, Tatiana V. Gani$^{2}$ and Vakhid A. Gani$^{3,4}$}

\address{$^1$Moscow Institute of Physics and Technology,
%Institutskiy per.\ 9,
Dolgoprudny, Moscow Region 141700, Russia
}

\address{$^2$Faculty of Mathematics, National Research University Higher School of Economics, Moscow 119048, Russia
}

\address{$^3$Department of Mathematics, National Research Nuclear University MEPhI\\ (Moscow Engineering Physics Institute),
%Kashirskoe shosse 31,
Moscow 115409, Russia}

\address{$^4$Theory Department, Institute for Theoretical and Experimental Physics\\ of National Research Centre ``Kurchatov Institute'', 
%Bolshaya Cheremushkinskaya str.\ 25,
Moscow 117218, Russia}

\ead{vagani@mephi.ru}

\begin{abstract}
We consider the change in the asymptotic behavior of solutions of the type of flat domain walls (i.e.\ kink solutions) in field-theoretic models with a real scalar field. We show that when the model is deformed by a bounded deforming function, the exponential asymptotics of the corresponding kink solutions remain exponential, while the power-law ones remain power-law. However, the parameters of these asymptotics, which are related to the wall thickness, can change.
\end{abstract}

\section{Introduction and motivation}

{\it Domain wall} in three-dimensional physical space is a boundary between {\it domains} (areas of space) with different properties \cite{Vachaspati.book.2006}. In particular, in a field-theoretic model with a real scalar field with potential having at least two minima ({\it vacua} of the model), domain wall separates regions with different vacuum values of the scalar field. Thus, domain wall is a transition region in which the field continuously changes from one vacuum value to another. In the direction perpendicular to the wall, the field dependence on the spatial coordinate can be described by a {\it kink solution} of the corresponding differential equation \cite{Vachaspati.book.2006,Shnir.book.2018,Manton.book.2004}. Domain walls arise in a variety of physical models. In particular, the appearance of domain walls and their subsequent collapse in the Early Universe could lead to the formation of primordial black holes \cite{GaKiRu}. Emphasize that the study of many properties of domain walls is reduced to the study of kink solutions of the corresponding field-theoretic models. In this context, the wall thickness is related to the rate of asymptotic approach of the field to the vacuum value when moving away from the wall deep into the domain.

This our study is at the intersection of two areas of research related to kinks. On the one hand, we study asymptotic properties of kink solutions. On the other hand, we apply the so-called {\it deformation procedure} \cite{Bazeia.PRD.2002,Bazeia.EPJC.2018,Blinov.arXiv.2020} and investigate how this affects on the kink asymptotics. In short, in this paper, we present some preliminary results for transformational properties of the kink asymptotics with respect to deformations of the field-theoretic model.

Note that this short letter is based on the material presented by Tatiana Gani at the conference ICPPA-2020.

\section{Deformation procedure and two types of deforming functions}

Consider a real scalar field $\varphi(x,t)$ in $(1+1)$-dimensional space-time. Assume that dynamics of the field is described by the Lagrangian density
\begin{equation}\label{eq:lagrangian}
\mathcal{L} = \frac{1}{2}\left(\frac{\partial \varphi}{\partial t}\right)^2-\frac{1}{2}\left(\frac{\partial \varphi}{\partial x}\right)^2-V(\varphi)
\end{equation}
with the potential $V(\varphi)$ having two or more denenerate minima. Without loss of generality, we assume that the potential is a non-negative function that vanishes at its minima. Such model can have kink solutions, see, e.g., \cite[Part I]{Shnir.book.2018}, \cite[Chap.~5]{Manton.book.2004}.

The essence of the deformation procedure is briefly as follows. Assume that we have a field-theoretic model with the potential $V^{(0)}(\varphi)$ and its known kink solution $\varphi_{\rm K}^{(0)}(x)$. Also assume that there is a strictly monotonically increasing function  $f(\varphi)$ called {\it deforming function}. Then we can introduce a new model with the potential $V^{(1)}(\varphi)$ and kink solution $\varphi_{\rm K}^{(1)}(x)$:
\begin{equation}\label{eq:new_model}
    V^{(1)}(\varphi) = \frac{V^{(0)}[f(\varphi)]}{[f^\prime(\varphi)]^2}
    \quad \mbox{and} \quad
    \varphi_{\rm K}^{(1)}(x) = f^{-1}[\varphi_{\rm K}^{(0)}(x)].
\end{equation}
In this paper, we briefly discuss how the asymptotic behavior of kink changes under deformations by two different types of deforming functions:
\begin{itemize}
    \item type 1 --- a strictly monotonic with a finite derivative;
    \item type 2 --- a function which has infinite derivative at the vacuum of the deformed model.
\end{itemize}
We study transformational properties of two different types of the kink's asymptotics: {\it exponential} and {\it power-law}. On the one hand, there is a vast variety of field-theoretic models having kinks with exponential asymptotic behavior, for example, $\varphi^4$ \cite{Campbell.PhysD.1983,Belova.PhysD.1988,Goodman.SIAM_JADS.2005,Dorey.JHEP.2017,Dorey.PLB.2018,Askari.CSF.2020}, $\varphi^6$ \cite{Dorey.PRL.2011,Weigel.JPCS.2014,Demirkaya.JHEP.2017,Romanczukiewicz.PLB.2017,Lima.JHEP.2019}, $\varphi^8$ \cite{GaLeLi,Gani.PRD.2020}, double sine-Gordon \cite{Gani.EPJC.2018.dsg,Gani.EPJC.2019.dsg}, etc. \cite{Bazeia.EPJC.2018,Belendryasova.JPCS.2019.logKinks,Peyrard.PhysD.1983.msG}. On the other hand, many models admit kink solutions with power-law asymptotics, see, e.g., \cite{Belendryasova.CNSNS.2019,Christov.PRD.2019,Christov.arXiv.2020,Khare.JPA.2019,Manton.JPA.2019,Campos.arXiv.2020,dOrnellas.JPC.2020}.

\section{Transformational properties of the kink's asymptotics}

First of all, recall that the kink can have exponential or power-law asymptotics, depending on the behavior of the potential $V^{(0)}(\varphi)$ near the vacuum $\varphi=\varphi_0^{}$ (assume that $\lim\limits_{x\to+\infty}\varphi_{\rm K}^{(0)}(x)=\varphi_0^{}$), for more details see, e.g., \cite{Christov.PRD.2019}. If
\begin{equation}\label{eq:polynomial_potential_0_approx}
    V^{(0)}(\varphi) \approx \frac{1}{2}\left(\varphi_0^{}-\varphi\right)^{2k} v(\varphi_0^{})
\end{equation}
then the asymptotics of $\varphi_{\rm K}^{(0)}(x)$ is exponential for $k=1$:
\begin{equation}\label{eq:exponential_asymptotics_0}
    \varphi_{\rm K}^{(0)}(x) \approx \varphi_0^{} - \exp\left[-\sqrt{v(\varphi_0^{})}\:x\right] \quad \mbox{at} \quad x\to+\infty;
\end{equation}
and power-law for $k>1$:
\begin{equation}\label{eq:kink_asymptotics_0}
    \varphi_{\rm K}^{(0)}(x) \approx \varphi_0^{} - \frac{A_k^{(0)}}{x^{1/(k-1)}} \quad \mbox{at} \quad x\to+\infty,
\end{equation}
where
\begin{equation}\label{eq:A_0}
     A_k^{(0)} = \left[\left(k-1\right)\sqrt{v(\varphi_0^{})}\right]^{1/(1-k)}.
\end{equation}

We have found that after deformation of the model \eqref{eq:polynomial_potential_0_approx} by deforming function $f(\varphi)$ of the first type, the asymptotics of the deformed kink will be the following. The exponential asymptotics \eqref{eq:exponential_asymptotics_0} changes to
\begin{equation}\label{eq:exponential_asymptotics_1}
    \varphi_{\rm K}^{(1)}(x) \approx \varphi_1^{} - \exp\left[-\sqrt{v(\varphi_0^{})}\:x\right] \quad \mbox{at} \quad x\to+\infty,
\end{equation}
while the power-law asymptotics \eqref{eq:kink_asymptotics_0} changes to
\begin{equation}\label{eq:kink_asymptotics_1}
    \varphi_{\rm K}^{(1)}(x) \approx \varphi_1^{} - \frac{A_k^{(1)}}{x^{1/(k-1)}} \quad \mbox{at} \quad x\to+\infty,
\end{equation}
where
\begin{equation}\label{eq:A_1}
     A_k^{(1)} = \frac{1}{f^\prime(\varphi_1^{})} \left[\left(k-1\right)
     \sqrt{v(\varphi_0^{})}\right]^{1/(1-k)},
\end{equation}
so that
\begin{equation}\label{eq:A_ratio}
    \frac{A_k^{(0)}}{A_k^{(1)}} = f^\prime(\varphi_1^{}).
\end{equation}
Here $\varphi_1^{}=f^{-1}(\varphi_0^{})$ is the corresponding vacuum of the deformed model.

In the case of deforming function of the second type
\begin{equation}\label{eq:approx_f}
    f(\varphi) \approx f(\varphi_1^{}) - B\left(\varphi_1^{}-\varphi\right)^\beta \quad \mbox{at} \quad \varphi \to \varphi_1^{} - 0,
\end{equation}
where $B>0$ and $0<\beta<1$ are constants, i.e.\ $f^\prime(\varphi)$ is unbounded at $\varphi \rightarrow \varphi_1^{}-0$, we obtain the following asymptotics of the deformed kink. The exponential asymptotics \eqref{eq:exponential_asymptotics_0} changes to
\begin{equation}\label{eq:f-deformed_kink_01}
    \varphi_{\rm K}^{(1)}(x) \approx \varphi_1^{} - \exp{\left[ -\frac{\sqrt{v(\varphi_0^{})}}{\beta}\:x\right]} \quad \mbox{at} \quad x \to +\infty,
\end{equation}
while the power-law asymptotics \eqref{eq:kink_asymptotics_0} changes to
\begin{equation}\label{eq:f-deformed_kink_1}
    \varphi_{\rm K}^{(1)}(x) \approx \varphi_1^{} - \frac{B_k^{(1)}}{x^{\frac{1}{\beta(k-1)}}} \quad \mbox{at} \quad x \to +\infty,
  \end{equation}
where
\begin{equation}
    B_k^{(1)} = \left[\left(k-1\right)B_{}^{k-1}\sqrt{v(\varphi_0^{})}\right]^{\frac{1}{\beta (1-k)}}.
\end{equation}

\section{Concluding remarks}

To summarize, we have found the following transformational properties of the kink's asymptotics under the deformation procedure.

A. For strictly monotonic deforming function with finite derivative:
\begin{itemize}
    \item the exponential asymptotics after deformation remains exponential with the same constant coefficient in front of $x$;
    \item the power-law asymptotics after deformation remains power-law with the same power; however, depending on $f^\prime(\varphi)$, the numerical coefficient can change, Eq.~\eqref{eq:A_ratio}.
\end{itemize}

B. If the derivative of the deforming function goes to infinity in the vacuum of the deformed model:
\begin{itemize}
    \item the exponential asymptotics after deformation remains exponential, however, the coefficient in front of $x$ increases;
    \item the power-law asymptotics after deformation remains power-law, however, the power of $x$ in the denominator increases;
    \item we can say that for both asymptotics the kink solution of the deformed model approaches the vacuum value faster.
\end{itemize}

A more detailed description of these and some other results can be found in preprint \cite{Blinov.arXiv.2020} and, we hope, will be published.

\section*{Acknowledgements}

V.A.G.\ acknowledges the Russian Foundation for Basic Research for their support under Grant No.\ 19-02-00930.

This work was also supported by the MEPhI Academic Excellence Project.


\section*{References}

\begin{thebibliography}{99}

\bibitem{Vachaspati.book.2006}
T.~Vachaspati,
\href{https://doi.org/10.1017/CBO9780511535192}{\it Kinks and Domain Walls: An Introduction to Classical and Quantum Solitons},
Cambridge University Press, Cambridge U.K. (2006).

\bibitem{Shnir.book.2018}
Y.~M.~Shnir,
\href{https://doi.org/10.1017/9781108555623}{\it Topological and Non-Topological Solitons in Scalar Field Theories},
Cambridge University Press, Cambridge U.K. (2018).

\bibitem{Manton.book.2004}
N.~Manton and P.~Sutcliffe,
\href{https://doi.org/10.1017/CBO9780511617034}{\it Topological Solitons},
Cambridge University Press, Cambridge U.K. (2004).

\bibitem{GaKiRu}
V.~A.~Gani, A.~A.~Kirillov, and S.~G.~Rubin,
{\it Classical transitions with the topological number changing in the early Universe},
\href{https://doi.org/10.1088/1475-7516/2018/04/042}{{JCAP} {\bf 04}, 042 (2018)}
[\href{https://arxiv.org/abs/1704.03688}{\tt arXiv:1704.03688}].

\bibitem{Bazeia.PRD.2002}
D.~Bazeia, L.~Losano, and J.~M.~C.~Malbouisson,
{\it Deformed defects},
\href{https://doi.org/10.1103/PhysRevD.66.101701}{{Phys.\ Rev.\ D} {\bf 66}, 101701 (2002)}
[\href{https://arxiv.org/abs/hep-th/0209027}{\tt arXiv:hep-th/0209027}].

\bibitem{Bazeia.EPJC.2018}
D.~Bazeia, E.~Belendryasova, and V.~A.~Gani,
{\it Scattering of kinks of the sinh-deformed $\varphi^4$ model},
\href{https://doi.org/10.1140/epjc/s10052-018-5815-z}{{Eur.\ Phys.\ J.\ C} {\bf 78}, 340 (2018)}
[\href{https://arxiv.org/abs/1710.04993}{\tt arXiv:1710.04993}].

\bibitem{Blinov.arXiv.2020}
P.~A.~Blinov, T.~V.~Gani, and V.~A.~Gani,
{\it Deformations of Kink Tails},
\href{https://arxiv.org/abs/2008.13159}{\tt arXiv:2008.13159} (2020).

\bibitem{Campbell.PhysD.1983}
D.~K.~Campbell, J.~F.~Schonfeld, and C.~A.~Wingate,
{\it Resonance structure in kink-antikink interactions in $\varphi^4$ theory},
\href{https://doi.org/10.1016/0167-2789(83)90289-0}{{Physica D} {\bf 9}, 1 (1983)}.

\bibitem{Belova.PhysD.1988}
T.~I.~Belova and A.~E.~Kudryavtsev,
{\it Quasi-periodic orbits in the scalar classical $\lambda\phi^4$ field theory},
\href{https://doi.org/10.1016/0167-2789(88)90085-1}{{Physica D} {\bf 32}, 18 (1988)}.

\bibitem{Goodman.SIAM_JADS.2005}
R.~H.~Goodman and R.~Haberman,
{\it Kink-Antikink Collisions in the $\phi^4$ Equation: The $n$-Bounce Resonance and the Separatrix Map},
\href{https://doi.org/10.1137/050632981}{{SIAM J.\ Appl.\ Dyn.\ Syst.} {\bf 4}, 1195 (2005)}.

\bibitem{Dorey.JHEP.2017}
P.~Dorey et al.,
{\it Boundary scattering in the $\phi^4$ model},
\href{https://doi.org/10.1007/JHEP05(2017)107}{{JHEP} {\bf 05}, 107 (2017)}
[\href{https://arxiv.org/abs/1508.02329}{\tt arXiv:1508.0232}].

\bibitem{Dorey.PLB.2018}
P.~Dorey and T.~Roma\'nczukiewicz,
{\it Resonant kink--antikink scattering through quasinormal modes},
\href{https://doi.org/10.1016/j.physletb.2018.02.003}{{Phys.\ Lett.\ B} {\bf 779}, 117 (2018)}
[\href{http://arxiv.org/abs/1712.10235}{\tt arXiv:1712.10235}].

\bibitem{Askari.CSF.2020}
A.~Askari et al.,
{\it Collision of $\phi^4$ kinks free of the Peierls-Nabarro barrier in the regime of strong discreteness},
\href{https://doi.org/10.1016/j.chaos.2020.109854}{{Chaos, Solitons and Fractals} {\bf 138}, 109854 (2020)}
[\href{https://arxiv.org/abs/1912.07953}{\tt arXiv:1912.07953}].

\bibitem{Dorey.PRL.2011}
P.~Dorey, K.~Mersh, T.~Roma\'nczukiewicz, and Ya.~Shnir,
{\it Kink-antikink collisions in the $\phi^6$ model},
\href{https://doi.org/10.1103/PhysRevLett.107.091602}{{Phys.\ Rev.\ Lett.} {\bf 107}, 091602 (2011)}
[\href{https://arxiv.org/abs/1101.5951}{\tt arXiv:1101.5951}].

\bibitem{Weigel.JPCS.2014}
H.~Weigel,
{\it Kink-Antikink Scattering in $\varphi^4$ and $\phi^6$ Models},
\href{https://doi.org/10.1088/1742-6596/482/1/012045}{{J.\ Phys.: Conf.\ Ser.} {\bf 482}, 012045 (2014)}
[\href{https://arxiv.org/abs/1309.6607}{\tt arXiv:1309.6607}]

\bibitem{Demirkaya.JHEP.2017}
A.~Demirkaya et al.,
{\it Kink dynamics in a parametric $\phi^6$ system: A model with controllably many internal modes},
\href{https://doi.org/10.1007/JHEP12(2017)071}{{JHEP} {\bf 12}, 071 (2017)}
[\href{http://arxiv.org/abs/1706.01193}{\tt arXiv:1706.01193}].

\bibitem{Romanczukiewicz.PLB.2017}
T.~Roma\'nczukiewicz,
{\it Could the primordial radiation be responsible for vanishing of topological defects?},
\href{https://doi.org/10.1016/j.physletb.2017.08.045}{{Phys.\ Lett.\ B} {\bf 773}, 295 (2017)}
[\href{https://arxiv.org/abs/1706.05192}{\tt arXiv:1706.05192}].

\bibitem{Lima.JHEP.2019}
F.~C.~Lima, F.~C.~Simas, K.~Z.~Nobrega, and A.~R.~Gomes,
{\it Boundary scattering in the $\phi^6$ model},
\href{https://doi.org/10.1007/JHEP10(2019)147}{{JHEP} {\bf 10}, 147 (2019)}
[\href{https://arxiv.org/abs/1808.06703}{\tt arXiv:1808.06703}].

\bibitem{GaLeLi}
V.~A.~Gani, V.~Lensky, and M.~A.~Lizunova,
{\it Kink excitation spectra in the $(1+1)$-dimensional $\varphi^8$ model},
\href{https://doi.org/10.1007/JHEP08(2015)147}{{JHEP} {\bf 08}, 147 (2015)}
[\href{https://arxiv.org/abs/1506.02313}{\tt arXiv:1506.02313}].

\bibitem{Gani.PRD.2020}
V.~A.~Gani, A.~Moradi~Marjaneh, and P.~A.~Blinov,
{\it Explicit kinks in higher-order field theories},
\href{https://doi.org/10.1103/PhysRevD.101.125017}{{Phys.\ Rev.\ D} {\bf 101}, 125017 (2020)}
[\href{https://arxiv.org/abs/2002.09981}{\tt arXiv:2002.09981}].

\bibitem{Gani.EPJC.2018.dsg}
V.~A.~Gani et al.,
{\it Scattering of the double sine-Gordon kinks},
\href{https://doi.org/10.1140/epjc/s10052-018-5813-1}{{Eur.\ Phys.\ J.\ C} {\bf 78}, 345 (2018)}
[\href{https://arxiv.org/abs/1711.01918}{\tt arXiv:1711.01918}].

\bibitem{Gani.EPJC.2019.dsg}
V.~A.~Gani, A.~Moradi~Marjaneh, and D.~Saadatmand,
{\it Multi-kink scattering in the double sine-Gordon model},
\href{https://doi.org/10.1140/epjc/s10052-019-7125-5}{{Eur.\ Phys.\ J.\ C} {\bf 79}, 620 (2019)}
[\href{https://arxiv.org/abs/1901.07966}{\tt arXiv:1901.07966}].

\bibitem{Belendryasova.JPCS.2019.logKinks}
E.~Belendryasova, V.~A.~Gani, and K.~G.~Zloshchastiev,
{\it Kinks in the relativistic model with logarithmic nonlinearity},
\href{https://doi.org/10.1088/1742-6596/1390/1/012082}{{J.\ Phys.: Conf.\ Ser.} {\bf 1390}, 012082 (2019)}
[\href{https://arxiv.org/abs/2001.02265}{\tt arXiv:2001.02265}].

\bibitem{Peyrard.PhysD.1983.msG}
M.~Peyrard and D.~K.~Campbell,
{\it Kink-antikink interactions in a modified sine-Gordon model},
\href{https://doi.org/10.1016/0167-2789(83)90290-7}{{Physica D} {\bf 9}, 33 (1989)}.

\bibitem{Belendryasova.CNSNS.2019}
E.~Belendryasova and V.~A.~Gani,
{\it Scattering of the $\varphi^8$ kinks with power-law asymptotics},
\href{https://doi.org/10.1016/j.cnsns.2018.07.030}{{Commun.\ Nonlinear Sci.\ Numer.\ Simulat.} {\bf 67}, 414 (2019)}
[\href{https://arxiv.org/abs/1708.00403}{\tt arXiv:1708.00403}].

\bibitem{Christov.PRD.2019}
I.~C.~Christov et al.,
{\it Long-range interactions of kinks},
\href{https://doi.org/10.1103/PhysRevD.99.016010}{{Phys.\ Rev.\ D} {\bf 99}, 016010 (2019)}
[\href{https://arxiv.org/abs/1810.03590}{\tt arXiv:1810.03590}].

\bibitem{Christov.arXiv.2020}
I.~C.~Christov et al.,
{\it Kink-Antikink Collisions and Multi-Bounce Resonance Windows in Higher-Order Field Theories},
\href{https://arxiv.org/abs/2005.00154}{\tt arXiv:2005.00154} (2020).

\bibitem{Khare.JPA.2019}
A.~Khare and A.~Saxena,
{\it Family of potentials with power law kink tails},
\href{https://doi.org/10.1088/1751-8121/ab30fd}{{J.\ Phys.\ A: Math.\ Theor.} {\bf 52}, 365401 (2019)}
[\href{https://arxiv.org/abs/1810.12907}{\tt arXiv:1810.12907}]

\bibitem{Manton.JPA.2019}
N.~S.~Manton,
{\it Forces between kinks and antikinks with long-range tails},
\href{https://doi.org/10.1088/1751-8121/aaf9d1}{{J.\ Phys.\ A: Math.\ Theor.} {\bf 52}, 065401 (2019)}
[\href{https://arxiv.org/abs/1810.03557}{\tt arXiv:1810.03557}].

\bibitem{Campos.arXiv.2020}
J.~G.~F.~Campos and A.~Mohammadi,
{\it Interaction between kinks and antikinks with double long-range tails},
\href{https://arxiv.org/abs/2006.01956}{\tt arXiv:2006.01956} (2020).

\bibitem{dOrnellas.JPC.2020}
P.~d'Ornellas,
{\it Forces between kinks in $\phi^8$ theory},
\href{https://doi.org/10.1088/2399-6528/ab90c2}{{J.\ Phys.\ Commun.} {\bf 4}, 055014 (2020)}
[\href{https://arxiv.org/abs/2001.10744}{\tt arXiv:2001.10744}].

\end{thebibliography}
\end{document}